\documentclass{aa}  
\usepackage[varg]{txfonts}

\usepackage{graphicx}
\usepackage{natbib}
\usepackage{color}
\usepackage{subfig}
\usepackage[switch]{lineno}
\usepackage{xspace}
\usepackage{microtype}
\usepackage{amssymb}
\usepackage{mathtools}
\usepackage{multirow}
\def \link_col{blue}
\usepackage{xspace}
\usepackage{url}
\usepackage{wasysym}
\usepackage{caption}
\usepackage{upgreek}

\usepackage{graphicx}   
\usepackage{dblfloatfix}
\usepackage{amsmath}    
\usepackage{threeparttable}
\usepackage{subcaption}
\usepackage[dvipsnames]{xcolor}
\usepackage[normalem]{ulem}
\usepackage{booktabs}

\def \link_col{blue}
\newcommand{\gray}{$\gamma$-ray~}

\newcommand{\highlight}[1]{{\color{black} #1}}

\makeatletter
\DeclareRobustCommand*{\fieldName}[1]{%
  \begingroup\@fieldName\scantokens{\texttt{\small {#1}}\noexpand}\endgroup}
\begingroup\lccode`\~=`\_\relax
   \lowercase{\endgroup\def\@fieldName{\catcode`\_=\active \let~\_}}
\makeatother

\begin{document}

\title{Nuclear de-excitation line emissions from giant molecular clouds}

\author{Zhaodong Shi\inst{1,2,3}
\and Bing Liu\inst{1,2,3}\fnmsep\thanks{Corresponding author; \texttt{lbing@ustc.edu.cn}}
\and Rui-zhi Yang\inst{1,2,3,4}
}
\institute{
CAS Key Laboratory for Research in Galaxies and Cosmology, Department of Astronomy, School of Physical Sciences, \\
University of Science and Technology of China, Hefei, Anhui 230026, China
\and Deep Space Exploration Laboratory, Hefei 230088, China
\and School of Astronomy and Space Science, University of Science and Technology of China, Hefei, Anhui 230026, China 
\and TIANFU Cosmic Ray Research Center, Chengdu, Sichuan, China
}
\date{Received 23 August 2024 / Accepted 6 November 2024}
\titlerunning{Nuclear de-excitation line emissions from giant molecular clouds}
\authorrunning{Shi et al.}
\abstract{
Understanding how cosmic rays (CRs) propagate within the giant molecular clouds (GMCs) is critical for studying the dynamics and chemical processes inside the clouds. The flux of low-energy CRs inside the dense cores of GMCs strongly affects the heating and ionization of the gases and further influences the star-forming process. We analytically calculated the CR distribution inside GMCs assuming different diffusion coefficients, and estimated the corresponding nuclear de-excitation line emission and the ionization rate resulting from the interaction between the penetrating CRs and gases. We find that future MeV observations can be used as a unique probe to measure the low-energy CR density in situ and test different CR propagation scenario inside GMCs.}

\maketitle

\section{Introduction}
\label{sec:intro}

Giant molecular clouds (GMCs) are made of cold molecular gases. Stars are believed to be formed in the dense cores of GMCs. On the one hand, the star formation processes are the ultimate energy source that accelerates cosmic rays (CRs), either through stellar winds or via supernova remnants and compact objects at later stages of stellar evolution. On the other hand, the CRs also play an important role in regulating the star formation process by governing the heating and ionization processes in the star-forming regions \citep{dalgarno06}, affecting the dynamical evolution of gas and initiating several crucial chemical reactions \citep{papadopoulos10}. The initial mass function can also be affected by the CR density \citep{papadopoulos10}. In addition, CRs are also likely responsible for the ionization of $\mathrm{H_2}$ molecules observed in the diffuse clouds in the central molecular zone \citep{dogiel2013}. Furthermore, CRs are believed to permeate GMCs and produce $\gamma$ rays through the inelastic scattering of CRs with nuclei in the gas. Thus, GMCs are also regarded as CR barometers \citep{aharonian01}. Therefore, the interplay between star formation in GMCs and CRs is regarded as one of the most important physical processes in astrophysics. 

However, the propagation of CRs inside GMCs is far from understood. The magnetic turbulence that scatters the CRs is believed to be damped out in the dense neutral gas environment in GMCs \citep{cesarsky78}; as such, the propagation inside these dense regions should be described as free streaming or advection. 
The faster transport leads to an effective diffusion coefficient that is larger than those in the interstellar medium \citep[ISM;][]{cesarsky78, morlino15}. In this case, CRs should penetrate freely into the dense core of GMCs. However, recent studies based on \highlight{\textit{Fermi} Large Area Telescope (\textit{Fermi}-LAT)} 
GeV observations of nearby GMCs show that CRs with energy lower than $10\, {\rm GeV}$ cannot penetrate the dense cores \citep{taurus}. Such a shielding of low-energy CRs (LECRs) can be explained by the slow diffusion of CRs inside GMCs. 
In quasi-linear theory, slower diffusion requires stronger magnetic turbulence, which can occur if the energy density of the magnetic field is a fixed fraction of that of the kinematic motion, which is further tied to the potential energy if the system is virialized. 
\citep{2016MNRAS.459.2432V,2017MNRAS.464.4096L,2018MNRAS.474.2167L}. Another possibility is the magnetic turbulence induced by CR streaming instability, resulting in slower diffusion inside dense clouds \citep{skilling1976, dogiel18}. Consequently, the diffusion property would also determine the CR distribution inside GMCs, which would have important implications for star processes and affect the resulting emissions of these objects. 

The predominance of free-streaming or self-generated turbulence over each other in the transport of CRs inside GMCs is highly relevant to the conditions and phases of GMCs and the surrounding ISM, which are very different in the cores of GMCs and their diffuse envelopes \citep{ivlev2018}. One crucial ingredient affecting the above processes is the generation of magnetic fields in GMCs, which is likely driven by turbulent motions 
\citep{istomin2013}. Moreover, a partial and weak ionization in GMCs has significant effects on the development of magnetohydrodynamic (MHD) waves and the cascade of MHD turbulence, such as via ion-neutral damping, and consequently on the transport of CRs \citep{xu2016}. In addition, depending on the configuration of magnetic fields, the CR flux inside GMCs can be reduced due to magnetic mirroring if it dominates over the magnetic focusing  \citep{silsbee2018, owen2021}. All these processes and their interplay can have important influences on the transport of CRs inside GMCs. However, it is very complicated to take all these processes into account, and we will not address these aspects in the present article. Instead, motivated by the recent \gray observations of nearby GMCs such as the Taurus and Perseus clouds \citep{taurus} and Orion B \citep{zeng2024}, which show a depletion of LECRs, we used the simple diffusion process to describe the transport of CRs inside GMCs and calculated the resulting MeV nuclear de-excitation \gray line emission. The nuclear \gray line emission by itself is not sensitive to the \highlight{aforementioned} processes and thus is a robust and powerful probe of the distribution of CRs within GMCs, independent of the ionization rate measurements, which are entangled with the environment of GMCs.


In this study we analytically calculated the CR distributions inside GMCs assuming different diffusion coefficients. In addition to the \gray emission from $\pi^0$-decay processes in the inelastic scattering of CRs with ambient gas, which is discussed in \citet{taurus},  we calculated the MeV de-excitation line emission \citep{Ramaty1979} and ionization rate of CRs, \highlight{which are mainly induced by LECRs with energies below the energy threshold ($\sim$ 280 MeV) of the $\pi^0$-decay processes that are generated by proton-proton (p-p) inelastic collisions.} 

This paper is organized as follows.
In Sect. \ref{sec:cr} we describe the methods used for the simulation of CR penetration in the GMCs and show the resulting CR spectra under different assumptions. In Sect. \ref{sec:nlines} we then calculate the corresponding nuclear de-excitation line emission, continuum \gray emission, and the 6.4 keV Fe K$\alpha$ line emission produced by the interaction of intrusion CRs with the molecular gases. 
We discuss the results in Sect. \ref{sec:dis}.

\section{CR distribution within the molecular clouds}
\label{sec:cr}
The transportation of CRs inside molecular clouds can be described phenomenologically as a diffusion process with energy losses taken into account \citep{gabici2007}. The energy losses are especially important to LECRs with energies $\lesssim 1\ \mathrm{GeV}$, though they only have minor influences on the higher-energy CRs. The advection and adiabatic loss can be safely ignored considering the turbulent bulk velocity is quite low in molecular clouds  ($\sim 10~\mathrm{km/s}$). Therefore, the transport equation of CRs inside molecular clouds is

\begin{equation} \label{eq:diffloss}
  \frac{\partial N}{\partial t} - \boldsymbol{\nabla}\cdot(D\boldsymbol{\nabla}N) - \frac{\partial }{\partial p}(bN) = Q,
\end{equation}

\noindent where $N(\mathbf{r},\,p,\,t) = 4\pi p^2 f(\mathbf{r},\,p,\,t)$ ($f$ is the isotropic distribution function of CRs) is the number density per momentum of CRs, $D(\mathbf{r},\,p)$ is the (isotropic) spatial diffusion coefficient, $b(\mathbf{r},\,p)=-dp/dt$ is the momentum loss rate, and $Q(\mathbf{r},\,p,\,t)$ is the source function of CRs. Generally, the gas density of molecular clouds is inhomogeneous, so the diffusion and energy losses are space-dependent.

The dominant energy loss process for CR nuclei is ionization interaction with the ambient gases \citep{schlickeiser2002}, and above the kinetic energy threshold $E_\mathrm{k,th} = 0.2787\ \mathrm{GeV}$, the inelastic nuclear p-p 
collision becomes significant. The lifetime of protons due to the inelastic p-p collision is $\tau_{\mathrm{pp}}=1/(\kappa n_\mathrm{gas} c\beta \sigma_\mathrm{pp})$, where the parameterized inelastic cross section, $\sigma_{\mathrm{pp}}$, is taken from \citet{kafexhiu2014}, the inelasticity $\kappa \approx 0.45$ \citep{aharonian1996}, $n_\mathrm{gas}$ is the gas density, and $\beta = v/c$ is the velocity of CRs in units of the speed of light, $c$. The energy loss resulting from the Coulomb collision with the ambient gases is very small compared with the above two processes for the typical parameter regime of molecular clouds \citep{padovani2009}.

\begin{figure}
\centering
\includegraphics[width=0.49\textwidth]{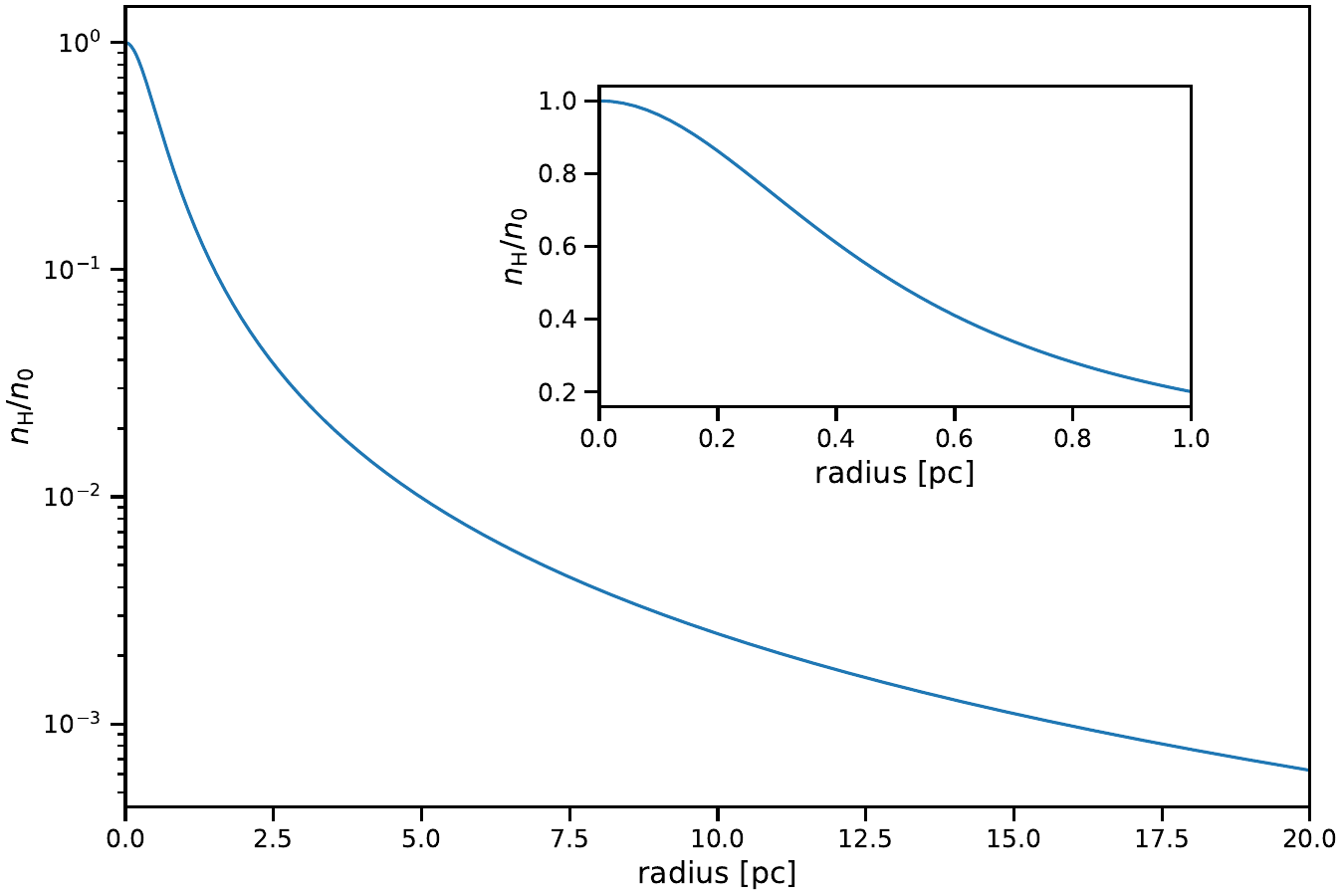}
\caption{Gas density radial profile of a 20 pc GMC with a dense core of radius $R_\mathrm{c}$ = 0.5 pc and profile index $\alpha=2$.}
\label{fig:gmc1_profile}
\end{figure}

For the sake of simplicity, we considered the clouds to be spherically symmetric. 
The gas density profile of the molecular cloud as a function of radius, $r,$ takes the following parameterization \citep{gabici2007}:
\begin{equation}
n_{\mathrm{H}}=n_{\mathrm{HI}} + 2n_{\mathrm{H}_2} = \frac{n_0}{1+(r/R_\mathrm{c})^{\alpha}},
\end{equation}

\noindent where $R_\mathrm{c}$ is the radius of the dense core, $n_0$ is the gas density at the center of the cloud, and $\alpha$ is the profile index. Figure \ref{fig:gmc1_profile} shows the gas density radial profile of a 20 pc GMC whose dense core radius $R_\mathrm{c}$ = 0.5 pc and profile index $\alpha=2$. The magnetic field profile is related to the gas density as \citep{crutcher2012,padovani2018}

\begin{equation}
B = 10 \left(\frac{n_\mathrm{H}}{300\ \mathrm{cm^{-3}}}\right)^{1/2}\ \mathrm{\upmu G}.
\end{equation}

\noindent The spatial diffusion coefficient is parameterized as

\begin{equation}
  D(p) = D_0 \beta \left( \frac{p / 1\ \mathrm{GeV} c^{-1}}{B / 3 \ \mathrm{\upmu G}}\right)^{\delta}.
\end{equation}

As the boundary condition for Eq. \eqref{eq:diffloss}, we assumed that the CR spectra at the boundary of the cloud are the local interstellar spectra (LIS). The parameterized best-fit LIS model for protons up to 100 GeV \highlight{obtained by \citet{vos2015}} according to recent experiment data, including from Voyager 1 and 2, \highlight{Payload for Antimatter Matter Exploration and Light-nuclei Astrophysics (PAMELA)}, 
and \highlight{Alpha Magnetic Spectrometer (AMS-02)}, is 

\begin{equation}
  j_{\mathrm{LIS}}(E_{\mathrm{k}}) = 2.70\frac{E_{\mathrm{k}}^{1.12}}{\beta^2}\left(\frac{E_{\mathrm{k}} + 0.67}{1.67}\right)^{-3.93} \ \mathrm{m^{-2} s^{-1} sr^{-1} MeV^{-1}},
\end{equation}

\noindent where $E_{\mathrm{k}}$ is the kinetic energy of CR protons (in GeV). The flux spectrum as a function of kinetic energy is $j(E_{\mathrm{k}}) = \frac{1}{4\pi} N(p)$.

The setting of this boundary condition is justified for the nearby GMCs since the recent \textit{Fermi}-LAT observations demonstrate that the spatial distribution of CRs in the Galaxy is homogeneous within a few parsecs of the Sun and the spectra are close to the CRs directly measured by AMS-02 and Voyager 1 and 2 \citep{yang14, neronov2017, aharonian2020}. Due to their proximity, it should be much easier for future MeV \gray telescopes to detect the nuclear de-excitation line emission generated by \highlight{the LECRs in} the nearby GMCs.

Eq. \eqref{eq:diffloss} has no analytical solution except when the gas distribution inside the cloud is uniform, and hence the numerical solution is needed. As a limiting case, however, we first give the analytical solution, which is also beneficial for verifying the numerical scheme adopted. 




Given that both the spatial diffusion and momentum loss are space-independent, the solution of Eq. \eqref{eq:diffloss} is 

\begin{equation} \label{eq:analytical-solution}
  \begin{aligned}
  N(r,\,p,\,t) &= N_0(p) + \\
  & \iiint dr'dp'dt'\, G(r,\,p,\,t;\,r',\,p',\,t') S(r',\,p',\,t'),
  \end{aligned}
\end{equation}

\noindent where the effective source term

\begin{equation}
  S(r,\,p,\,t)= Q(r,\,p,\,t) + \frac{\partial [b(p)N_0(p)]}{\partial p},
\end{equation}

\noindent and Green's function $G(r,\,p,\,t;\,r',\,p',\,t')$ satisfies

\begin{equation}
  \frac{\partial G}{\partial t} - \frac{1}{r^2}\frac{\partial}{\partial r}\left(r^2D\frac{\partial G}{\partial r}\right) - \frac{\partial (bG)}{\partial p} = \delta(r-r',\,p-p',\,t-t')
\end{equation}

\noindent and is subject to the boundary condition $G(R,\,p,\,t;\,r',\,p',\,t')=0,$ with $R$ the dimension of the cloud. Expanding into Fourier series, we have

\begin{equation}
\begin{aligned}
  &G(r,\,p,\,t;\,r',\,p',\,t') \\ &= \frac{1}{b(p)}\frac{1}{2\pi R}\sum_{m=1}^\infty \left\{  4\pi {r'}^2\frac{\sin(m\pi r'/R)}{r'} \frac{\sin(m\pi r/R)}{r} \right.\\ & \left. \exp\left[ -\frac{m^2\pi^2}{R^2}\lambda(p,\,p') \right] \theta(\lambda(p,\,p'))\delta(t-t'-\tau(p,\,p')) \right\},
\end{aligned}
\end{equation}

\noindent where $\theta(x)$ is the Heaviside step function, and we have introduced \citep{syrovatskii1959}

\begin{equation}
\tau(p,\,p')= \int_{p}^{p'}\frac{dp_1}{b(p_1)},\quad \lambda(p,\,p')=\int_{p}^{p'}\frac{D(p_1)}{b(p_1)}dp_1.
\end{equation}

\begin{figure}
\centering
\includegraphics[width=0.45\textwidth]{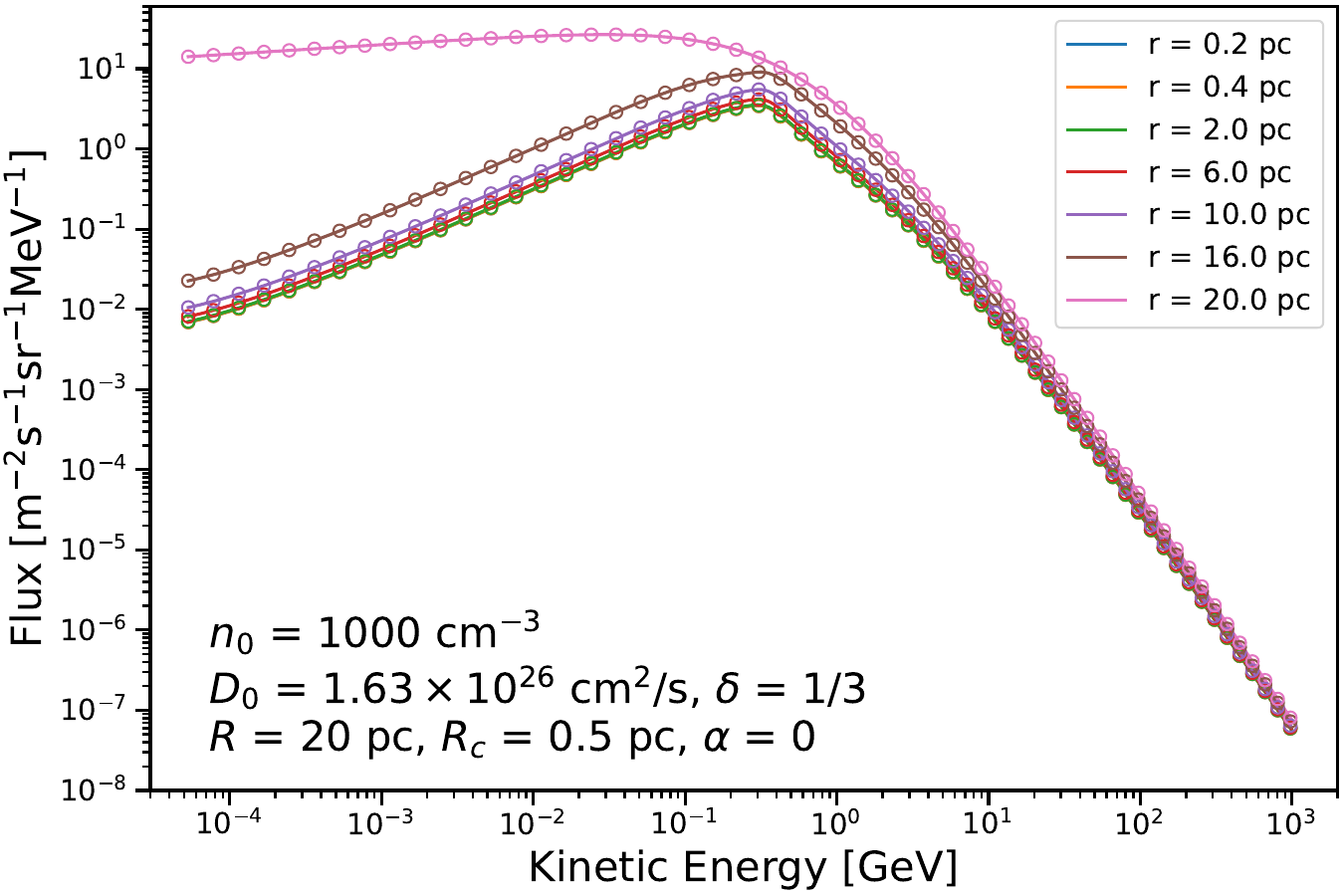}
\caption{Numerical solutions (solid lines) and the analytical solutions according to Eq. \eqref{eq:analytical-solution} (open circles). The time step $\Delta t$ = 20 years for numerical solutions in this case. Here, we have assumed that the molecular cloud is passive (i.e., there are no CR sources inside it) and homogeneous ($\alpha$ = 0) with a constant gas density of $n_\mathrm{H} = 500\ \mathrm{cm^{-3}}$ ($n_0=1000\ \mathrm{cm^{-3}}$ due to $x^0=1$). We have also assumed that the abundance of gas helium is 0.0925 relative to the gas H density, i.e., $n_{\mathrm{He}}=0.0925n_\mathrm{H}$. The dimension of the cloud $R$ = 20 pc. The spatial diffusion coefficient $D_0 = 1.63 \times 10^{26}\ \mathrm{cm^2/s,}$ and the exponent $\delta=1/3$. }
\label{fig:verif}
\end{figure}

To numerically solve Eq. \eqref{eq:diffloss}, we adopted the operator splitting method, used the implicit upwind scheme for the momentum loss term, and used the Crank-Nicolson scheme for the diffusion term 
\citep{nr1992, hanasz2021}. As a verification of the numerical schemes, Fig. \ref{fig:verif} shows a comparison of the numerical solutions (solid lines) with the analytical solutions (open circles) assuming that the cloud is homogeneous with a constant gas density $n_\mathrm{H}=500\ \mathrm{cm}^{-3}$ -- which amounts to a gas H mass of $M_\mathrm{H} = \int m_\mathrm{H} n_\mathrm{H} dV \approx 4 \times 10^5\,{\rm M}_{\odot}$ (solar mass) and is typical of GMCs -- and there are no CR sources inside it. 
The numerical and analytical solutions match very well.

In the present study, we mainly considered the cases in which the clouds are ``passive'', that is to say, there are no CR sources inside them and they are only illuminated by incoming CRs from the outside ISM. Cases in which the clouds are ``active'' (i.e., there are CR sources embedded in GMCs) should be addressed separately. Due to the possible self-confinement of CRs around their sources, such as supernova remnants \citep{jacobs2022}, the CRs can be more intense inside active GMCs than in the local ISM \citep{baghmanyan2020}.

For the transport of CRs inside GMCs, the diffusion coefficient and the mass of GMCs are the two particularly crucial parameters. For the latter, we considered the two cases with gas masses of $M_\mathrm{H} = 10^5~{\rm M}_{\odot}$ and $M_\mathrm{H} = 10^6~{\rm M}_{\odot}$, respectively. The diffusion coefficient of CRs in the ISM is about $10^{28}\ \mathrm{cm^2/s}$ at 1 GeV based on investigations of CR propagation in the Galaxy \citep{strong2007}. However, recent \gray observations of nearby molecular clouds demonstrate that the transport of CRs inside them is much slower than in the ISM \citep{taurus}. Therefore, we used two different diffusion coefficients, $D_0 = 4 \times 10^{26} \ \mathrm{cm^2/s}$ and $D_0 = 4 \times 10^{28}\ \mathrm{cm^2/s}$, for each GMC mass  under consideration. The diffusion exponent, $\delta$, is poorly constrained by the current CR data due to a lack of \highlight{detailed knowledge} about the interstellar turbulence \citep{silver2024}. By taking three different values of the exponent (i.e., $\delta$ = 0.3, 0.5, and 0.7), we checked the effects of different $\delta$ on the CR transport for each pair of $D_0$ and $M_\mathrm{H}$. 

In addition, we assumed that the dimension (radius) of the GMCs is $R = 20 \ \mathrm{pc}$, which is typical of the Galactic GMCs according to observational statistics \citep{miville2017}, and that the radius of dense cores is $R_\mathrm{c} = 0.5 \ \mathrm{pc}$, which is consistent with observations \citep{draine2011}. As for the gas density radial profile, we assumed that the profile index is $\alpha=2$. While the profile is flat in the center (see Fig. \ref{fig:gmc1_profile}), consistent with observations \citep{bergin2007}, it is proportional to $r^{-2}$ at large radii, $r$, which results from the scale-free gravitational collapse \citep{ligx2018}. 

\begin{figure}
\centering
\includegraphics[width=0.45\textwidth]{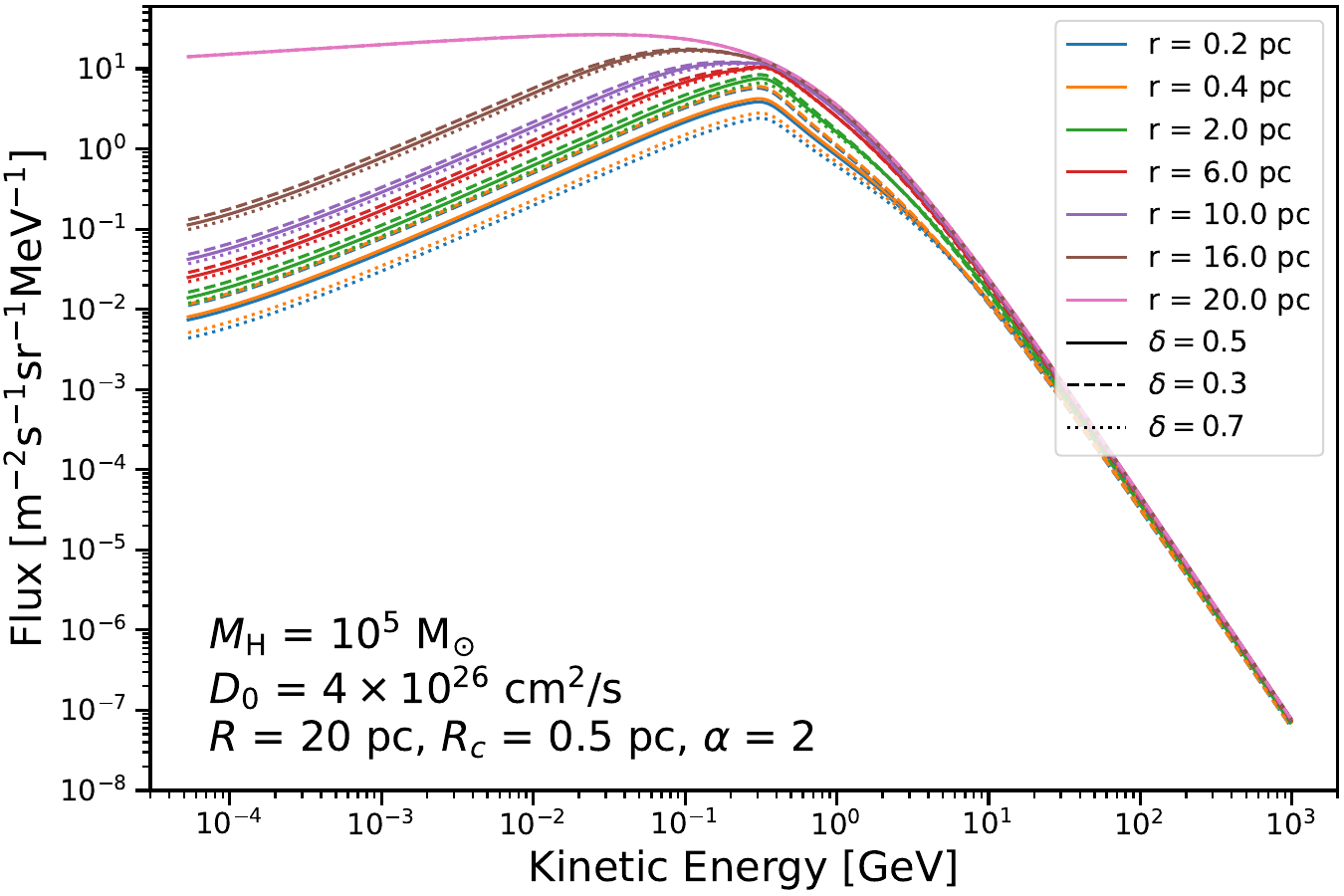}
\includegraphics[width=0.45\textwidth]{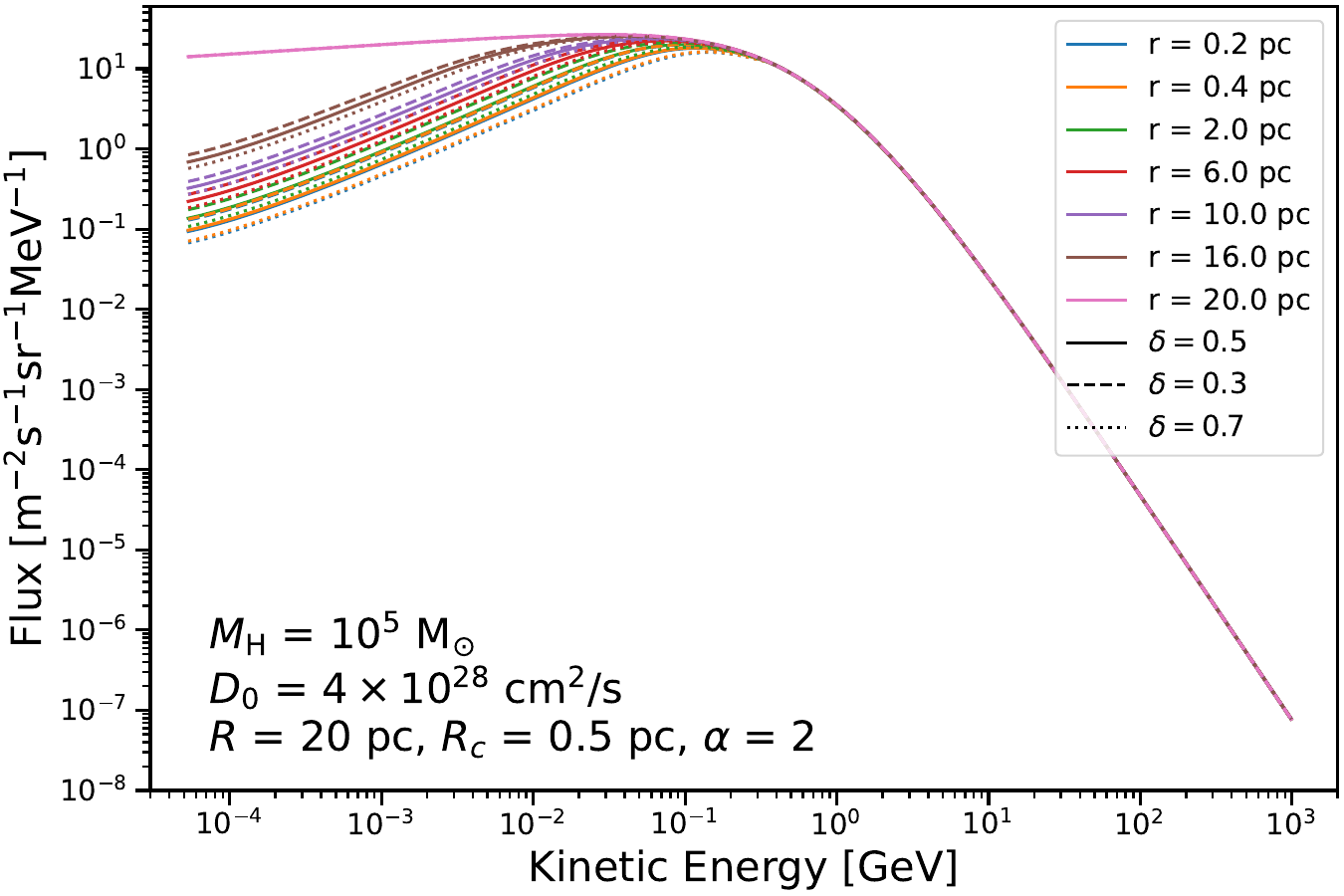}
\caption{
CR fluxes in a GMC with a gas H mass of $M_\mathrm{H}=10^5\ {\rm M}_{\odot}$.
Top panel: CR fluxes at different radii penetrating into a GMC with a dimension of 20 pc, assuming there are no sources inside it, for three different diffusion exponents: $\delta$ = 0.5 (solid lines), 0.3 (dashed lines), and 0.7 (dotted lines). The spatial diffusion coefficient $D_0=4\times 10^{26}~\mathrm{cm^2/s}$. The gas density profile index $\alpha=2$, and the core radius $R_\mathrm{c}$ = 0.5 pc. A fraction of the gas helium, whose abundance relative to H is $n_{\mathrm{He}}/n_{\mathrm{H}}=0.0925$, is also included. Bottom panel: Same as the top panel, but the spatial diffusion coefficient $D_0=4\times 10^{28}\ \mathrm{cm^2/s}$. }
\label{fig:gmc1}
\end{figure}

\begin{figure}
\centering
\includegraphics[width=0.45\textwidth]{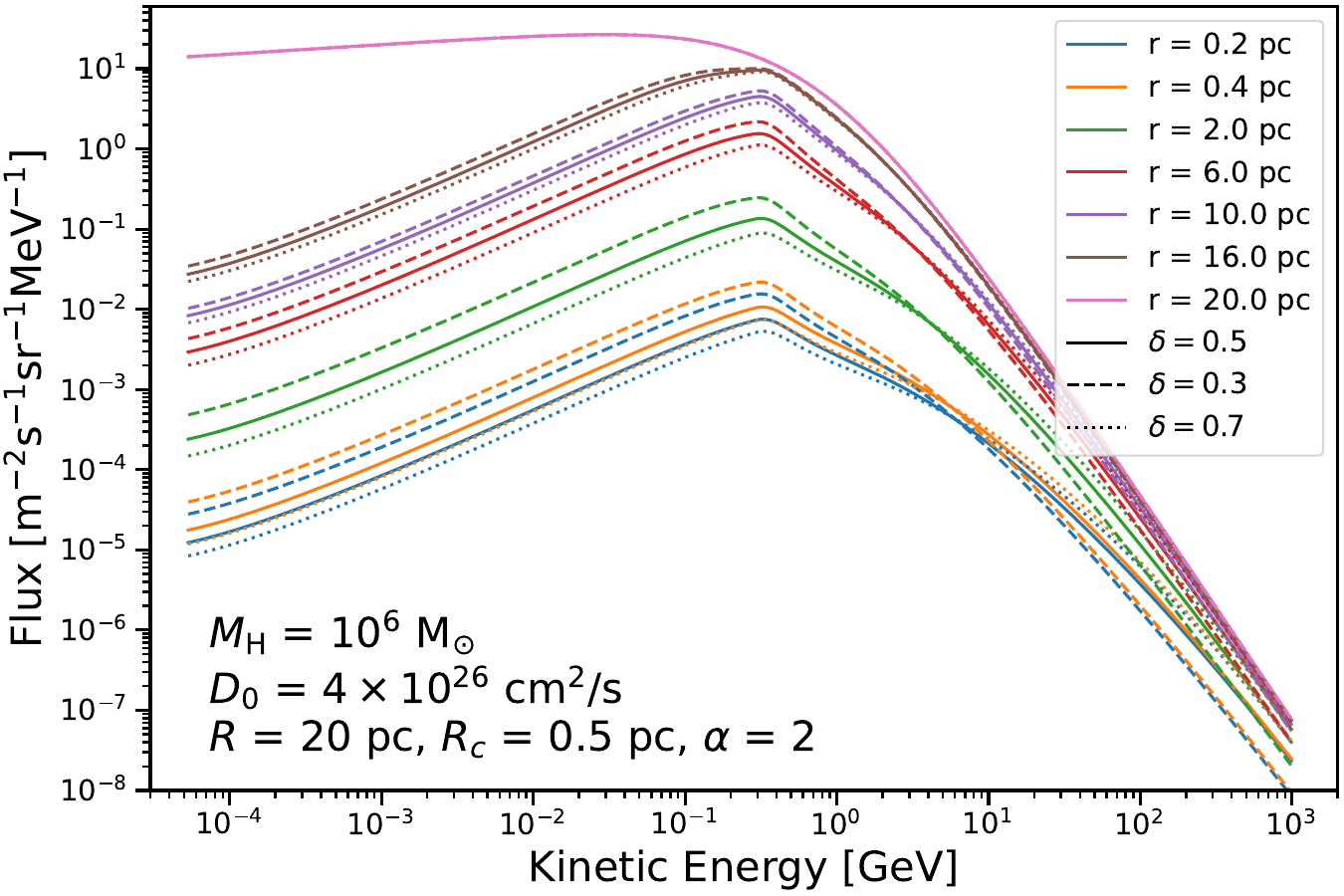}
\includegraphics[width=0.45\textwidth]{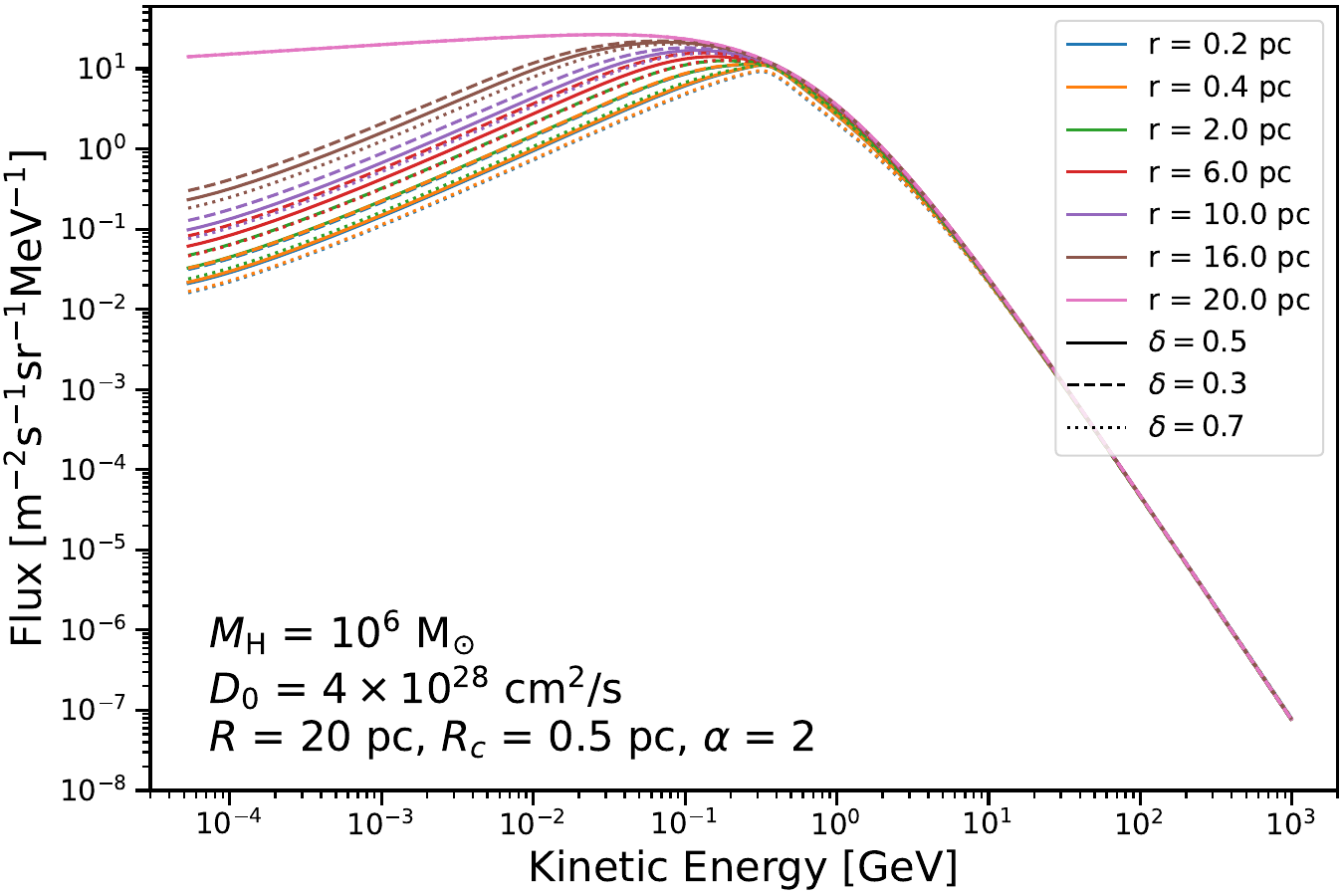}
\caption{Same as Fig. \ref{fig:gmc1}, but the gas H mass $M_\mathrm{H}=10^6\ {\rm M}_{\odot}$.}
\label{fig:gmc2}
\end{figure}

Figure \ref{fig:gmc1} shows the CR fluxes at different radii for $D_0 = 4 \times 10^{26} \ \mathrm{cm^2/s}$ (top panel) and $D_0 = 4 \times 10^{28} \ \mathrm{cm^2/s}$ (bottom panel) within a GMC with a mass of $M_\mathrm{H} = 10^5~{\rm M}_{\odot}$. In both cases, the intensities of CRs below 100 MeV are attenuated significantly, and such an attenuation is expected to have profound effects on the MeV nuclear de-excitation line emission. Additionally, for the case with a slower diffusion (top panel), the intensities of CRs below 10 GeV are also attenuated appreciably in the deepest interior of the GMC.
Similarly, Fig. \ref{fig:gmc2} shows the CR fluxes at different radii for $D_0 = 4 \times 10^{26} \ \mathrm{cm^2/s}$ (top panel) and $D_0 = 4 \times 10^{28} \ \mathrm{cm^2/s}$ (bottom panel) within a GMC with a mass of $M_\mathrm{H} = 10^6~{\rm M}_{\odot}$. For the slower diffusion (top panel), the intensities of CRs are attenuated very significantly even up to 1000 GeV, showing that CRs can just barely penetrate the dense core of such a massive GMC. 
In contrast, for the case with a faster diffusion (bottom panel), the CR intensities are only attenuated significantly below about 100 MeV.
In addition, as shown in Figs. \ref{fig:gmc1} and \ref{fig:gmc2},  an increase in the $\delta$ value, which will cause a further decrease in the diffusion coefficient, $D(p),$ of LECRs with $p<1\,{\rm GeV}c^{-1}$, consequently lowers their fluxes, especially at small radii of the massive clouds ($M_\mathrm{H}=10^6~\mathrm{M}_{\odot}$). Thus, the fluxes of the MeV de-excitation line emission that is mainly produced by the interaction among the LECR nuclei and the gases will also decrease due to the increase in the $\delta$ value. 
However, the change in LECR fluxes due to the shift in $\delta$ (from 0.3 to 0.7) is less than one order of magnitude, not as significant as that due to the variation in $D_0$ (from $4\times 10^{28} \ \mathrm{cm^2/s}$ to $4\times 10^{26} \ \mathrm{cm^2/s}$), which is multiple orders of magnitude. 
Therefore, we focus on the influence of $D_0$ and $M_\mathrm{H}$ in the following calculations and use a fiducial $\delta = 0.5$, corresponding to an Iroshnikov-Kraichnan MHD turbulence spectrum, for the sake of simplicity. 

\section{MeV nuclear de-excitation line emission and other observables}
\label{sec:nlines}

With the CR spectral and spatial distribution calculated above, we first estimated the possible MeV nuclear de-excitation line emission resulting from the nuclear interaction between the CR nuclei and the molecular gases, applying the method developed by \citet{Ramaty1979} and \citet{Murphy2009}.  We used numerical simulation results provided by the TALYS code \citep{talys2008, Koning2014} when the line cross sections adopted from the laboratory measurements of \citet{Murphy2009} and \citet{Benhabiles2013} were inadequate. 
As a general assumption, we chose the local values measured by Voyager \citep[see][their Table~3]{Cummings2016} for the elemental composition of the injected CRs and the recommended present-day solar abundances \citep[see][their Table~6]{Lodders2010} for the molecular cloud. 
For simplicity, we only considered two processes during the calculation: (i) the  CR protons and $\alpha$ particles excite heavier elements of the ambient gas, and (ii) the hydrogen and helium of the ambient gas collide with the heavy nuclei of CRs, in which only abundant elements, such as C, N, O, Ne, Mg, Si, S, Ar, Ca, and Fe, are taken into account.

\begin{figure}
\centering
\includegraphics[width=0.45\textwidth]{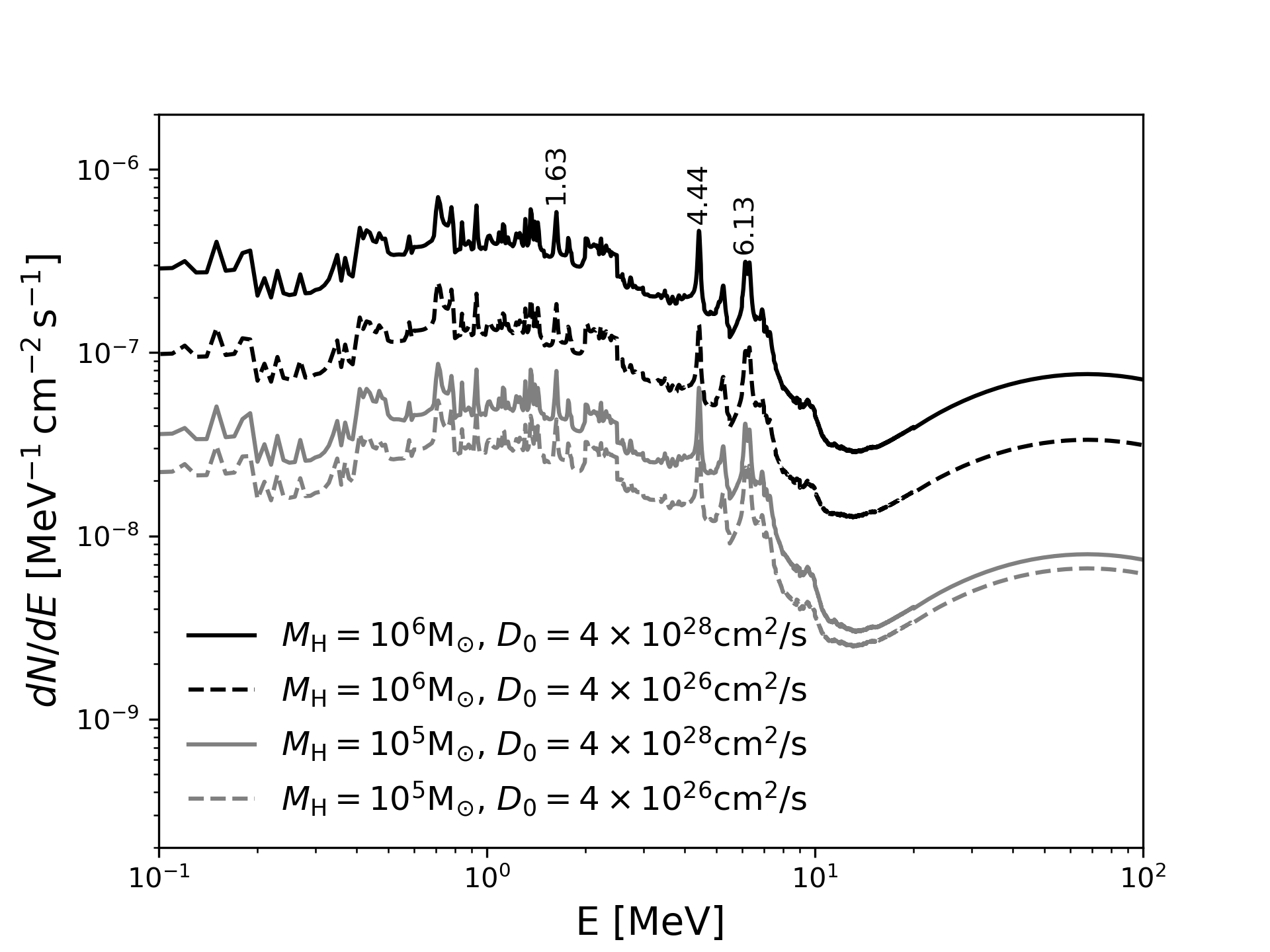}
\caption{ MeV nuclear de-excitation line emission and $\pi^0$-decay \gray emission fluxes integrated over entire GMCs of different masses assuming a distance of 200 pc. 
The diffusion coefficient of the CRs is $D_0=4\times 10^{28}\ \mathrm{cm^2/s}$ (solid lines) or $D_0=4\times 10^{26}\ \mathrm{cm^2/s}$  (dashed lines), and the mass of the molecular cloud is $1\times10^{6}\,{\rm M}_{\odot}$ (black lines) or $1\times10^{5}\,{\rm M}_{\odot}$ (gray lines). }
\label{fig:nlines_pion_gmc}
\end{figure}

\begin{figure}
    \centering
    \includegraphics[width=0.48\textwidth]{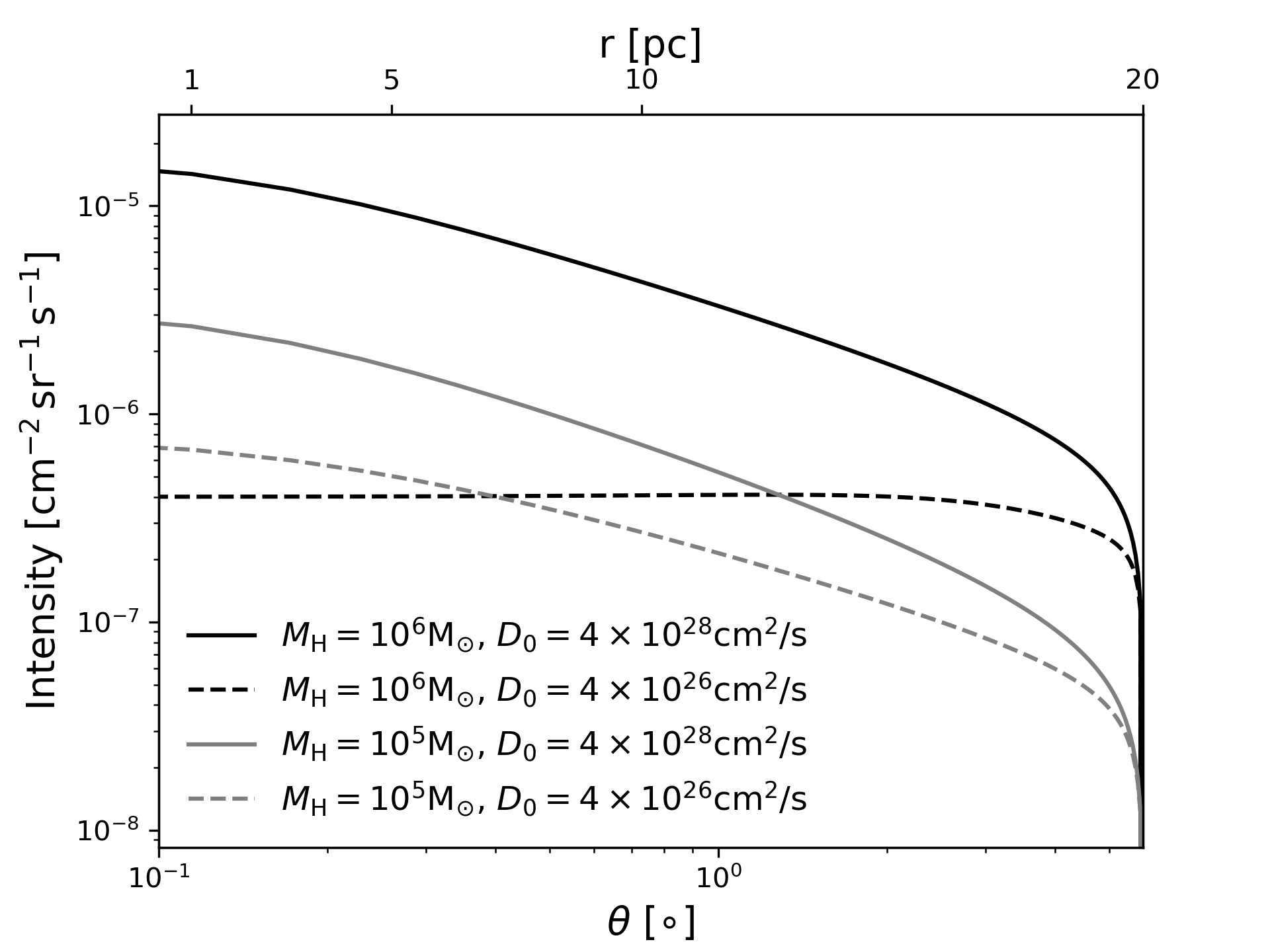}
   \caption{Integrated flux of strong narrow line emission at 4.44 MeV as a function of angular distance, $\theta$ (or radial distance, $r$) from the hypothetical molecular clouds located at a distance of 200\,pc.  The diffusion coefficient of the CRs is $D_0=4\times 10^{28}\ \mathrm{cm^2/s}$ (solid lines) or $D_0=4\times 10^{26}\ \mathrm{cm^2/s}$  (dashed lines), and the mass of the molecular cloud is $1\times10^{6}\,{\rm M}_{\odot}$ (black lines) or $1\times10^{5}\,{\rm M}_{\odot}$ (gray lines). }
    \label{fig:nlines_r}
    
\end{figure}

In addition to the nuclear de-excitation line emission,  the interaction between the penetrating CR nuclei and gases in the clouds will also inevitably generate nonthermal X-rays, such as the Fe K$\alpha$ line at 6.4 keV, via the collisional ionization of Fe, and continuum $\pi^0$-decay $\gamma$ rays via inelastic p-p collisions. Therefore, we also calculated the corresponding 6.4 keV Fe K$\alpha$ line emission using cross sections given by \citet{Tatischeff2012} and $\pi^0$-decay $\gamma$ rays by applying parameterized production cross sections from \citet{kafexhiu2014}. The differential fluxes of the nuclear de-excitation line emission and the $\pi^0$-decay \gray emission from CRs interacting with the cloud for different diffusion coefficients and total masses are shown in Fig. \ref{fig:nlines_pion_gmc}. The integrated fluxes of major narrow lines at 4.44 MeV and 6.13 MeV, which are mainly emitted via the de-excitation of $^{12}$C and $^{16}$O, and the 6.4-keV Fe K$\alpha$ line emission are presented in Table \ref{tab:flux}. %
If we only consider the intrusion of CRs from the local ISM, the fluxes of the corresponding \gray emissions induced by the interactions of penetrating CRs and the clouds are proportional to the total masses of the clouds.  Slower diffusion will yield lower fluxes for the same cloud, especially when the cloud is much more massive. In addition, we investigated the 2D distribution of these line emissions and illustrate the integrated flux of strong narrow line emission at 4.44 MeV as a function of angular distance, $\theta,$ in Fig. \ref{fig:nlines_r}. Here we assume the GMC is located about 200 pc away from Earth. Thus, the angular radius of a GMC with a radius of 20 pc is about 6 degrees. We can see that slower diffusion and denser gases, which cause the LECRs to lose more energy as they penetrate the cloud, consequently flatten the profile of the surface intensity of the 4.44-MeV line emission. However, the impact of cloud densities on the intensity profile can be ignored if the diffusion coefficient of CRs is large enough. 

\begin{figure}
    \centering
    \includegraphics[width=0.48\textwidth]{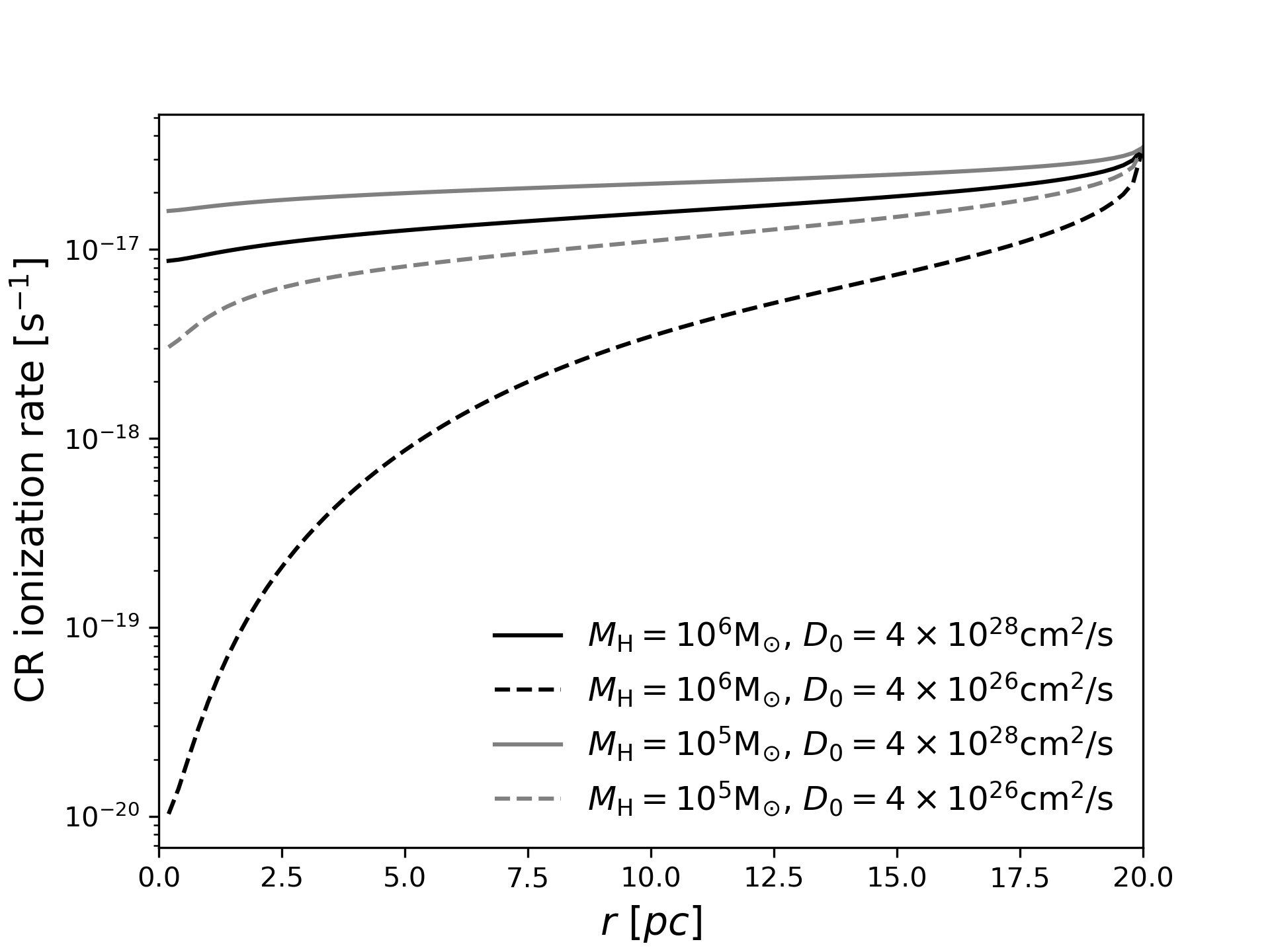}
   \caption{CR ionization rates at different distances from the center of the hypothetical GMCs. The diffusion coefficient of the CRs is $D_0=4\times 10^{28}\ \mathrm{cm^2/s}$ (solid lines) or $D_0=4\times 10^{26}\ \mathrm{cm^2/s}$  (dashed lines), and the mass of the molecular cloud is $1\times10^{6}\,{\rm M}_{\odot}$ (black lines) or $1\times10^{5}\,{\rm M}_{\odot}$ (gray lines). }
    \label{fig:ion_r}
\end{figure}

Moreover, the intrusive CRs will ionize the neutral gases in the clouds, and the CR ionization rate inside the GMC is strongly affected by the fluxes of LECRs.
Given the distribution of CRs calculated in Sect. \ref{sec:cr}, we calculated the CR ionization rate of molecular hydrogen by applying the formulae from \citet{Indriolo2009}. As shown in Fig. \ref{fig:ion_r}, if the diffusion coefficient, $D_0$, of the CRs within the cloud is the same as in the ISM, which is usually assumed to be $4\times 10^{28}\ \mathrm{cm^2/s}$, the variation in CR ionization rates between the interior and the outer layers of the cloud is not very large: the CR ionization rate within the dense core is about half that of the outer layers.
However, if the diffusion of CRs within the cloud is much slower, $D_0=4\times 10^{26}$, the CR ionization rate is about one magnitude lower at the center of a less massive cloud ($M_{\rm H}=1\times10^{5}\,{\rm M}_{\odot}$) compared to that at the outer boundary ($r=20$\,pc), and for more massive and denser clouds ($M_{\rm H}=1\times10^{6}\,{\rm M}_{\odot}$), the CR ionization rate can decrease by up to \highlight{three} orders of magnitude (from $3\times10^{-17} {\rm s}^{-1}$ at the boundary to $10^{-20} {\rm s}^{-1}$ at the center). 
To summarize, our results show that the closer to the center of the cloud, the lower the CR ionization rate, and they are consistent with previous theoretical calculations that predict the decrease in the CR ionization rate for increasing column density, $N({\rm H}_2)$ \citep[e.g.,][]{padovani2009}.

\begin{table}
\caption{Total
line fluxes obtained under different mass and diffusion coefficient assumptions.}
\begin{tabular}{ccccc}
\hline
$M_{\rm H}$ &$D_0$ & \multicolumn{3}{c}{Fluxes ($10^{-9}$\,ph\,cm$^{-2}$\,s$^{-1}$)}\\
 \cline{3-5}
 (${\rm M}_{\odot}$)&(${\rm cm}^{2}/{\rm s}$)&6.4 keV&4.44 MeV& 6.13 MeV\\
\hline
\multirow{2}{*}{1.0e+6}&$4\times10^{28}$&14.0&3.14&2.31\\
\cline{3-5}
&$4\times10^{26}$&4.94&0.98&0.75\\
 \hline
\multirow{2}{*}{1.0e+5}&$4\times10^{28}$&1.70&0.43&0.30\\
\cline{3-5}
&$4\times10^{26}$&1.10&0.23&0.17\\
\hline
\end{tabular}
\label{tab:flux}
\end{table}

\section{Discussion}
\label{sec:dis}

Recent analyses of GeV \gray emissions from \textit{Fermi}-LAT data, particularly in regions like the Taurus and Perseus molecular clouds, have revealed a potential slow diffusion of CRs within these clouds \citep{taurus}. This slow diffusion, along with the effective shielding of LECRs below 10 GeV from dense molecular clumps, has significant implications for the star formation process within these clumps. LECRs, particularly those with energies below 100 MeV, play a crucial role in gas ionization and heating, which are essential for regulating star formation. However, current observations of GeV-continuum $\gamma$ rays cannot provide information about CRs in this critical energy range.  The 6.4-keV line emissions have been detected from X-ray observations toward clouds, such as observations of a small region of $\rho$ Ophiuchi, the star-forming region of the Taurus cloud, and Sgr B, located in the Galactic Center region \citep[e.g.,][]{Imanishi2001,Gudel2007,Koyama2007}.  However, the origin of the detected X-ray point sources in the clouds is more likely related to stellar activities; for example, the \highlight{class I source (YLW 16A) in $\rho$ Ophiuchi} is associated with a young stellar object. 
Meanwhile, the origin of the diffuse Fe K$\alpha$ line emissions is more complex since X-ray photons and CR electrons can also induce such line emission.  In particular, for the 6.4 keV line emissions detected from the clouds very close to the Galactic Center, their flux, $\sim (10^{-7}$--$10^{-6})$\,ph\,cm$^{-2}$\,s$^{-1}$, is higher than what we estimated from the penetrating LECR nuclei in passive clouds, and their origin is still debatable \citep[e.g.,][]{Tatischeff2012, Dogiel2014}. Similarly, the constraints from the ionization rate on the properties of CRs in the clouds are also limited.
This makes MeV nuclear de-excitation line emissions particularly valuable, as they can directly probe LECR  nuclei and provide unique insights into their effects on the ISM. 

In this context, we investigated the 2D distribution of MeV line emissions from ideal GMCs relatively close to Earth. Specifically, we calculated the surface intensity of prominent narrow lines as a function of angular distance by integrating the line flux along the line of sight. For a GMC located 200 pc from Earth with a radius of 20 pc, the angular distance from the cloud's edge to its core is approximately 6 degrees. Using the 4.44 MeV line emission as an example, our findings indicate that in a massive cloud (with a hydrogen mass of $10^6\,{\rm M}_{\odot}$), the surface intensity of the emission line appears flattened due to the slow diffusion of CRs. This flattening suggests a strong shielding of LECRs within the cloud. As shown in Fig. \ref{fig:nlines_r}, the intensity of the 4.44 MeV line emission at the cloud's center is almost the same as that in its outer envelope (within 10 to 20 pc of the center). However, in less massive or less compact clouds, this flattening effect is less pronounced, and the center intensity of the 4.44 MeV line remains significantly higher than at the cloud's edge due to faster CR penetration or weaker shielding effects.
These results assume that no local CR sources are present within the GMCs. However, if CR acceleration occurs within the clouds — such as through star-forming processes — the distribution of CRs, and consequently the MeV line emission, will be different. Previous work has shown that CR distribution and MeV line emissions around a hypothetical accelerator that continuously injects particles into a uniform medium can be significantly enhanced in regions with slower CR diffusion and denser environments compared to passive clouds \citep{nlines}.

In recent years, several MeV detector projects,  for example \highlight{All-sky Medium-Energy Gamma-ray Observatory eXplorer \citep[AMEGO-X;][]{amego-x}} 
and \highlight{Compton Spectrometer and Imager \citep[COSI;][]{cosi}}, 
have been proposed or are in development, meaning detailed studies of LECR-induced MeV line emissions will be more feasible in the near future. The MeV line sensitivities of these new-generation instruments are on the level of $10^{-6}$\,ph\,cm$^{-2}$\,s$^{-1}$ ($3\sigma$ detection in $10^6$\,s), which is about one magnitude better than current operating detectors, such as the SPectrometer on board \highlight{INTErnational Gamma-Ray Astrophysics Laboratory \citep[INTEGRAL;][]{integral}}, 
whose line sensitivities are on the level of  $10^{-5}$\,ph\,cm$^{-2}$\,s$^{-1}$. According to our estimation, MeV line emissions from  passive GMCs, with fluxes on the order of or below $10^{-8}$\,ph\,cm$^{-2}$\,s$^{-1}$ (assuming the cloud is 200 pc away), will be challenging to detect even with next-generation MeV telescopes despite their improved sensitivity. Thus, for future MeV observations, these MeV line emissions from clouds are more likely to be detected from stacking analyses of multiple nearby GMCs rather than research of a specific cloud.  Moreover, regions that contain massive gases and strong star-forming activities,  such as the central molecular zone in our Galaxy, are also promising targets for investigating the properties of LECRs via MeV line emissions \citep{cmz_nline}.

Nonetheless, the MeV de-excitation lines can provide in situ measurements of LECRs in dense clumps, offering a crucial tool for understanding the initial conditions of the star formation process. If LECR accelerators, such as young protostars \citep{Padovani2015}, are embedded within the clouds, the corresponding ionization rates and MeV line fluxes could be higher than in passive clouds. As a result, nuclear de-excitation line emissions are more likely to be detected from GMCs with ongoing star-forming activities, providing direct insights into the role of LECRs in these critical processes and improving our understanding of the early stages of star formation.

\section*{Data availability}

To calculate emissivities of the de-excitation \gray line lines,  we used the code TALYS \citep[version 1.96,][]{talys2008}, which can be downloaded from {\url{https://tendl.web.psi.ch/tendl_2019/talys.html}}. For a better match with the experiment data, we modified the deformation files of $^{14}$N, $^{20}$Ne, and $^{28}$Si using the results of \citet{Benhabiles2011}. 
We also used the production cross sections of the specific lines listed in the compilation of \citet{Murphy2009}.

\begin{acknowledgements}

Bing Liu acknowledges the support from the NSFC under grant 12103049. Rui-zhi Yang is supported by the NSFC under grants 12041305, and 12393854. The authors gratefully acknowledge the support of Cyrus Chun Ying Tang Foundations.

\end{acknowledgements}



\bibliographystyle{aa}
\bibliography{cite} 










\end{document}